\documentclass[pre,twocolumn,superscriptaddress,showpacs]{revtex4-1} 

\usepackage{amsmath}
\usepackage{amsfonts}
\usepackage{amssymb}
\usepackage[pdftex]{graphicx}
\usepackage{verbatim}
\usepackage{color}

\usepackage{bbm}

\newcommand{\indicatorsub}[1]{\mathbbm{1}_{#1}}

\usepackage[usenames,dvipsnames]{xcolor} 
\newcommand{\ER}{Erd\H{o}s-R\'{e}nyi }
\newcommand{\fracEasy}{\mathcal{E}}

\begin{document}
\title{Threshold cascades with response heterogeneity in multiplex networks}

\author{Kyu-Min Lee}
\affiliation{Department of Physics and Institute of Basic Science, Korea University, Seoul 136-713, Korea}
\author{Charles D. Brummitt}
\affiliation{Department of Mathematics and Complexity Sciences Center, University of California, Davis, CA 95616}
\author{K.-I. Goh}
\email{kgoh@korea.ac.kr}
\affiliation{Department of Physics and Institute of Basic Science, Korea University, Seoul 136-713, Korea}

\date{\today}

\begin{abstract}
Threshold cascade models have been used to describe spread of behavior in social networks and cascades of default in financial networks.  In some cases, these networks may have multiple kinds of interactions, such as distinct types of social ties or distinct types of financial liabilities; furthermore, nodes may respond in different ways to influence from their neighbors of multiple types. To start to capture such settings in a stylized way, we generalize a threshold cascade model to a multiplex network in which nodes follow one of two response rules:
some nodes activate when, in at least one layer, a large enough fraction of neighbors are active, while the other nodes activate when, in all layers, a large enough fraction of neighbors are active. Varying the fractions of nodes following either rule facilitates or inhibits cascades. Near the inhibition regime, global cascades appear discontinuously as the network density increases; however, the cascade grows more slowly over time. This behavior suggests a way in which various collective phenomena in the real world could appear abruptly yet slowly.
\end{abstract}

\maketitle

\section{Introduction} 
Multiple channels of interaction in a network (or network layers) can have nontrivial consequences in the system's dynamics and function~\cite{Buldyrev2010, Gao2011, LeichtDSouza2009, KMLee2012, Brummitt2012, Cozzo2012, Gomez2013, Coevolution,Coevolution2,NoN,multilayer_review,Viability,Azimi-Tafreshi2014}. 
Effects of introducing new network layers include catastrophic cascades of failure~\cite{Buldyrev2010}, facilitated cascades~\cite{Brummitt2012}, and a super-diffusive state~\cite{Gomez2013}.
Most such studies on multiplex networks have assumed identical dynamics for every node. 
However, many real-world complex systems 
consist of heterogeneous agents who could respond differently to their multiplex environment. For example, an individual engaging in a multiplex social network may respond to her social influences coming from multiple layers, and how she integrates those layer-level influences may depend on the individual's personal and communal characteristics. The consequences of such heterogeneous response to the layer-level influences on dynamic processes on multiplex networks have yet to be characterized and understood. 

The threshold cascade model has provided a theoretical tool for understanding the spread of behavior in a social network~\cite{Schelling73, Granovetter, Watts2002, Gleeson2007} and for studying  the cascades of ``knock-on'' default among financial institutions~\cite{Allen2000, Gai2010, May2010,  Battiston2012b}. In this stylized model, nodes exist in one of two states, active or inactive (e.g., a person changed behavior or not, a bank has defaulted or not). Initially, each node draws a threshold from a distribution $Q(r)$. An inactive node with degree $k$ and with $m$ active neighbors activates when its fraction of active neighbors, $m / k$, exceeds its threshold $r$. The dynamics are iterated starting from a small fraction $\rho_{0}$ of initially active ``seed'' nodes, and then the cascade size $\rho$, the fraction of active nodes in the steady state, is observed. Previous studies showed that, for a wide range of network densities and threshold distributions, even an extremely small seed fraction $\rho_0$ can activate a finite fraction of an infinite-size network, an event called a global cascade~\cite{Watts2002,Gleeson2007}. Recently, some studies have generalized the threshold model to temporal networks~\cite{Karimi} and to multiplex networks~\cite{Yagan, Brummitt2012}. However, cascades with heterogeneous nodal responses to their multiplex environments are not yet addressed. On this basis, in this paper we explore the effects of heterogeneous responses of nodes to their multiple layers in threshold cascade dynamics on multiplex networks by generalizing the previous work \cite{Brummitt2012}. 

This paper is organized as follows. In section II, we describe the model. The analytical approach for calculating the cascade size for multiplex networks with locally-treelike layers is presented in section III. In section IV, we discuss results obtained from the analytical calculations introduced in section III and from numerical simulations. Section V concludes with a discussion of these results and of ways to make the model more realistic.

\begin{figure*}[t]
\centering
\includegraphics[width=.85\textwidth]{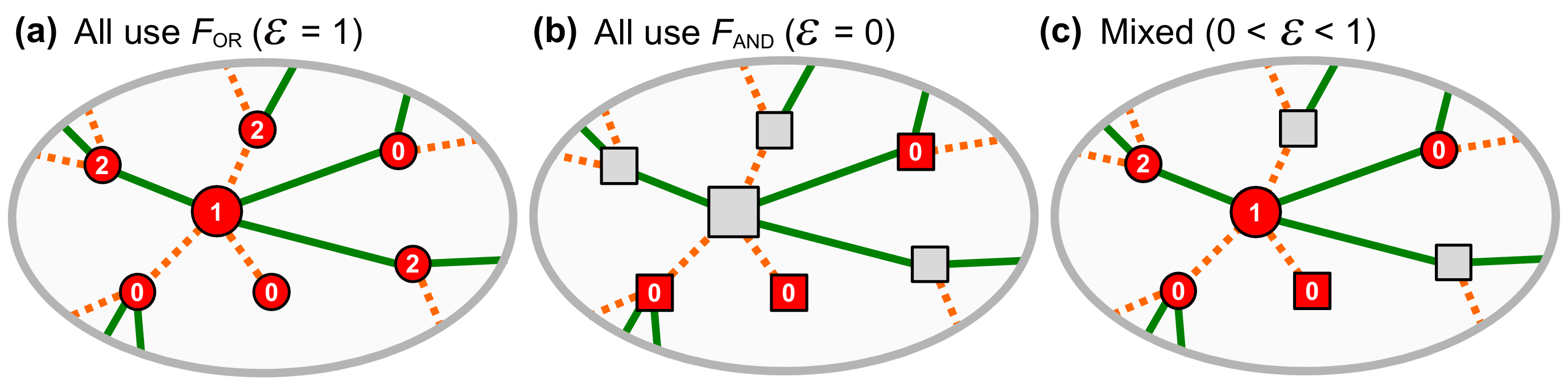}
\caption{(Color online) Illustration of the threshold model with heterogeneous responses on a two-layer multiplex network. Edges of the two layers are drawn as green (solid) and orange (dashed) lines, respectively. 
Circle-shaped nodes use the OR rule, while square-shaped nodes use the AND rule. Three cases for different fractions of OR nodes $\fracEasy$ are shown: (a) all nodes follow the OR rule $(\fracEasy=1)$; (b) all nodes follow the AND rule $(\fracEasy=0)$; and (c) nodes following either rule are mixed together $(0<\fracEasy<1)$. The numeric label on a node denotes the step at which the node activates, starting from the initial active seeds (labeled ``$0$''). Unlabeled nodes are those remaining inactive.
}
\label{fig:model_illustration}
\end{figure*}

\section{Model}
To keep theoretical simplicity and analytical tractability, we consider a mixture of populations of two types of nodes with simplified response rules---hereafter called ``OR nodes'' and ``AND nodes''. OR nodes activate as soon as, in at least one layer, a sufficiently large fraction of their neighbors in that layer are active. AND nodes are more stubborn: they activate as soon as, in each and every layer, a sufficiently large fraction of their neighbors in that layer are active. Figure~\ref{fig:model_illustration} depicts examples of cascades in small networks containing (a) all OR nodes and (b) all AND nodes.

Although highly stylized, the two types of response rules can be motivated by real-world multiplex system dynamics. In social systems, for instance, the OR rule would mean that just one social sphere can convince someone to change behavior, whereas the AND rule would mean that a person waits to change behavior until receiving enough influence from all social spheres. As another example, in banking systems, the OR rule would mean that a bank engaging in multiple kinds of lending defaults if, for at least one type of lending, sufficiently many of its borrowers of that type defaulted and cannot repay the bank, whereas the AND rule would mean that a bank defaults once enough of its borrowers of every type have defaulted. 

The two response rules have opposite effects. If all nodes respond with the OR rule as depicted in Fig.~\ref{fig:model_illustration}(a), then the existence of multiple layers facilitates global cascades~\cite{Brummitt2012}. On the contrary, if more nodes follow the AND rule [as illustrated in Fig.~\ref{fig:model_illustration}(b) and~\ref{fig:model_illustration}(c)], then global cascades  become rare or even impossible. As the system approaches this extreme of inhibited cascades,  global cascades appear discontinuously as the network densifies. We show that this phenomenon is associated with a cusp catastrophe and that it suggests ways to promote or to inhibit cascading phenomena in multiplex networks.

\section{Analytical approach}
\subsection{Mean cascade size $\rho$}
To investigate threshold dynamics on multiplex networks with heterogeneous layer responses, we first present the analytical approach for calculating the mean cascade size, applicable to multiplex networks with sparse, locally-treelike layers.  The analytic approach developed for the threshold cascade model on single-layer networks~\cite{Gleeson2007} can be generalized to the case of multiplex networks with $\ell$-layers (``$\ell$-plex networks'') \cite{Brummitt2012} by following a mean-field-type reasoning similar to other models on multiplex networks~\cite{Buldyrev2010,Viability,Azimi-Tafreshi2014}.

The expected size $\rho$ of a cascade begun from a fraction $\rho_0$ of initially active seed nodes (chosen uniformly at random) can be approximated for locally treelike networks as 
\begin{align}
\label{rho_equation}
\rho = \rho_0 + (1-\rho_0) \sum_{\mathbf{k}=\mathbf{0}}^\infty P(\mathbf{k}) \sum_{\mathbf{m}=\mathbf{0}}^{\mathbf{k}} \prod_{\alpha = 1}^\ell B_{m_{\alpha}}^{k_{\alpha}}(q_{\infty}^{(\alpha)}) \bar{F}(\mathbf{m},\mathbf{k}).
\end{align}
Here, we approximate the (locally treelike) graph as a tree, and $q_{\infty}^{(\nu)}$ is the (limiting) probability that a node is activated by its children, given that its parent in layer $\nu$ is inactive. $P(\mathbf{k})$ is the joint degree distribution of the $\ell$-plex network, with degree vector $\mathbf{k} \equiv (k_1, \dots, k_\ell)$. 
The sum $\sum_{\mathbf{m} = \mathbf{0}}^{\mathbf{k}}$ runs over all $\ell$-component vectors $\mathbf{m}$, representing the numbers of active neighbors in each layer, and thus having entries $m_\alpha \in \{0, \dots, k_\alpha\}$ for each layer index $\alpha \in \{1, \dots, \ell\}$.  
$B_m^k(p)$ is shorthand notation for the binomial probability distribution, $\binom{k}{m} p^m (1-p)^{k-m}$. 
$\bar{F}(\mathbf{m},\mathbf{k})$ is the mean response function, the probability that a node with degree $\mathbf{k}$ and $\mathbf{m}$ active neighbors activates, averaged over all thresholds (described in more detail below). 
Equation~\eqref{rho_equation} gives the probability that a randomly chosen node is either a seed node (with probability $\rho_0$, given by the first term on the righthand side) or is not a seed node but is activated by its active neighbors through the response function $\bar{F}$ [given by the second term on the righthand side of Eq.~\eqref{rho_equation}].

The probabilities $\{q_{\infty}^{(\alpha)}: 1\le \nu\le\ell\}$ in Eq.~(1) are obtained as   
the fixed point of coupled recursion equations that are written vectorially as relations, written vectorially as  
\begin{align}
\label{recursion}
\mathbf{q}_{n+1}= \mathbf{g}(\mathbf{q}_{n}),
\end{align}
with the $\alpha$-th component of~\eqref{recursion} given by 
\begin{align}
\label{q_equations}
&q_{n+1}^{(\alpha)} = g^{(\alpha)}(\mathbf{q}_{n}) \equiv \rho_0 + (1-\rho_0) \sum_{\mathbf{k}=\mathbf{0}}^{\mathbf{\infty}} \frac{k_{\alpha} P(\mathbf{k})}{z_{\alpha}} \times \notag \\
& \sum_{m_{\alpha}=0}^{k_{\alpha}-1}\sum_{\{m_{\nu}\}=\mathbf{0}, \nu \neq \alpha}^{\{k_{\nu}\}} \!\!\!\! B_{m_{\alpha}}^{k_{\alpha}-1}(q_{n}^{(\alpha)})\prod_{\nu \neq \alpha}  B_{m_{\nu}}^{k_{\nu}}(q_{n}^{(\nu)}) \bar{F}(\mathbf{m},\mathbf{k}),
\end{align} 
starting from $q_0^{(\alpha)}=\rho_0$ for all $\alpha\in\{1,\dots,\ell\}$. Here, $q_{n}^{(\alpha)}$ is the probability  that a node located $n$ hops from the leaves of the tree is activated by its children given that its parent in layer $\alpha$ is inactive. The leaves of the tree are initially active with probability $q_{0}^{(\alpha)} \equiv \rho_0$.

\subsection{The response functions}
The mean response function $\bar{F}$ used in Eqs.~(1--3) is defined by 
\begin{align}
\label{response_function}
\bar{F}(\mathbf{m},\mathbf{k}) \equiv \int \! Q(\mathbf{r}) F(\mathbf{m}, \mathbf{k}, \mathbf{r}) \, d\mathbf{r},
\end{align}
with the assumption that nodes independently draw their thresholds $\mathbf{r} \in [0,1]^\ell$ from a distribution $Q(\mathbf{r})$. The response function $F(\mathbf{m}, \mathbf{k}, \mathbf{r})$ is the probability that a randomly-chosen node with degree $\mathbf{k}$, among which $\mathbf{m}$ are active neighbors, and threshold $\mathbf{r}$ becomes active. Next we introduce a response function that has heterogeneous layer response rules. 

A node in an $\ell$-plex network has $k_\alpha$ neighbors in each layer $\alpha \in \{1, \dots, \ell\}$. At a certain point in time, this node sees that $m_\alpha$ out of its $k_\alpha$ neighbors in layer $\alpha$ are active, and the node responds according to one of the two elementary response rules, $F_{\rm OR}$ or $F_{\rm AND}$, defined as follows. $F_{\rm OR}$ denotes the ``OR rule'', for which an inactive node activates when, in at least one layer $\alpha$ in which it has neighbors (i.e., in which $k_\alpha > 0$), the fraction of active neighbors, $m_{\alpha}/k_{\alpha}$, exceeds its threshold $r_{\alpha}$.  [Recall the 
example with a small network shown in Fig.~\ref{fig:model_illustration}(a).] This OR rule can be implemented in the response function as
\begin{align} \label{FOR}
F_{\rm OR}(\mathbf{m}, \mathbf{k}, \mathbf{r}) = 
\max \left \{ \indicatorsub{\frac{m_{\alpha}}{k_{\alpha}} > r_{\alpha}} \! : \! 1\leq \alpha\leq \ell, k_\alpha > 0 \right \}, 
\end{align}
where $\indicatorsub{A}$ is the indicator function ($\indicatorsub{A}=1$ if $A$ is true, else $\indicatorsub{A}=0$). 
Likewise, the response function of the ``AND rule'', for which nodes activate when enough neighbors in  every layer are active [recall Fig.~\ref{fig:model_illustration}(b)], can be written as 
\begin{align} \label{FAND}
F_{\rm AND}(\mathbf{m}, \mathbf{k}, \mathbf{r})=
\min \left \{ \indicatorsub{\frac{m_{\alpha}}{k_{\alpha}} > r_{\alpha}} \! :  \! 1\leq \alpha\leq \ell, k_\alpha > 0 \right \}
\end{align}
[i.e., replace $\max$ with $\min$ in~\eqref{FOR}].

In our model, a random fraction $\fracEasy$ (respectively, $1-\fracEasy$) of nodes in the network follow the OR (AND) rule,  called the OR (AND) nodes [Fig.~\ref{fig:model_illustration}(c)]. Then, $F$ can be taken to be the additive mixture of two elementary response functions [Eqs.~\eqref{FOR} and~\eqref{FAND}] parametrized by $\fracEasy \in [0,1]$,
\begin{align}\label{Fmix}
F(\mathbf{m}, \mathbf{k}, \mathbf{r}) = \fracEasy F_{\rm OR}(\mathbf{m}, \mathbf{k}, \mathbf{r}) + (1-\fracEasy) F_{\rm AND}(\mathbf{m}, \mathbf{k}, \mathbf{r}).
\end{align}

\begin{figure}[t]
\centering
\includegraphics[width=.93\columnwidth]{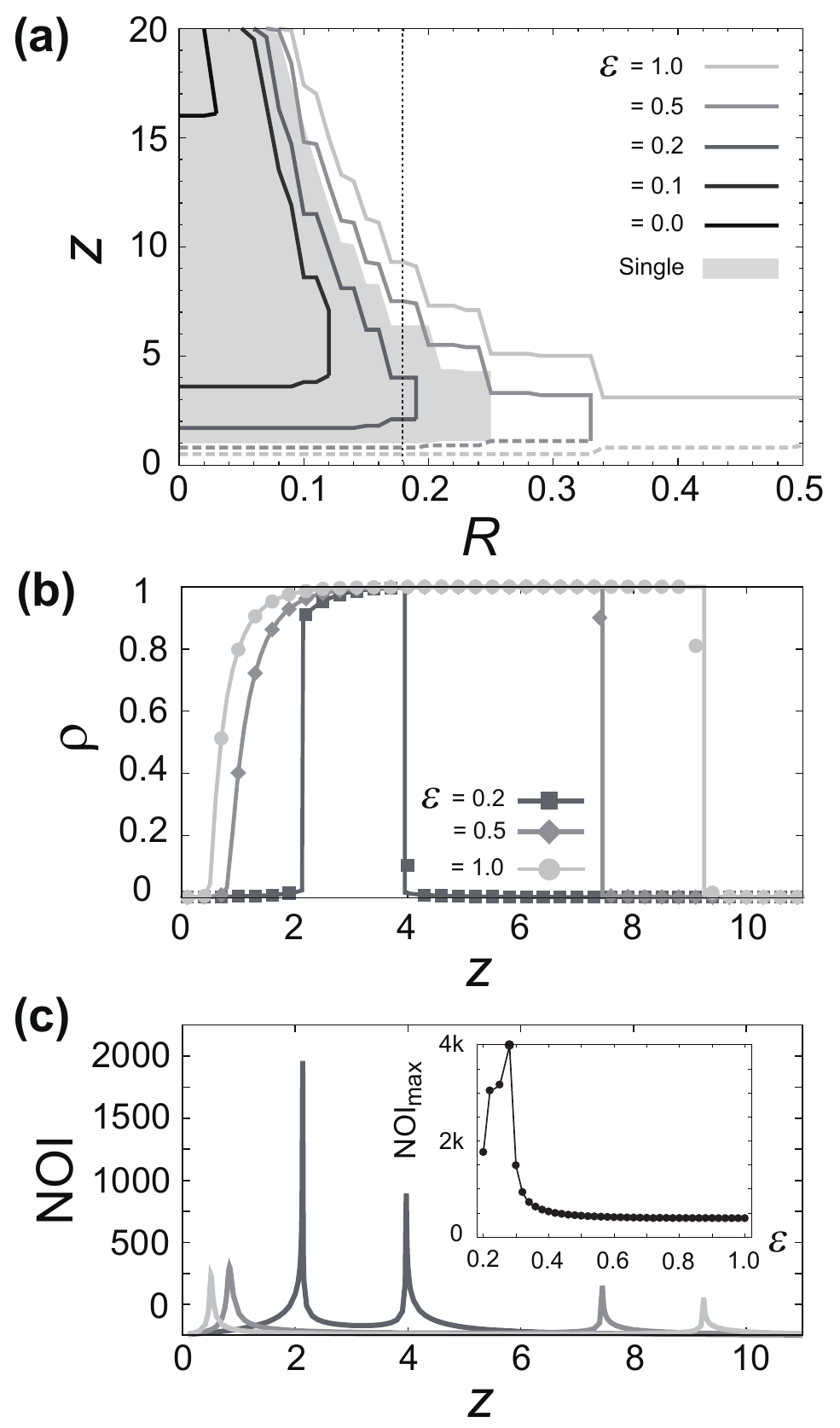} 
\caption{(a) The $(z,R)$-phase diagram of global cascades starting from $\rho_0 = 10^{-3}$ on duplex ER networks for various fractions $\fracEasy$ of OR nodes (lines) and for a single-layer network (gray shaded region). 
Global cascades occur on the left of the boundaries, which are obtained by finding local maxima of the number of iterations (NOI) of the recursion~\eqref{recursion} iterated until successive iterates differ by less than $10^{-10}$ (step sizes $\Delta z = 0.1, \Delta R = 0.01$). The dashed and solid lines indicate the continuous and discontinuous transitions, respectively. (b) Cascade size $\rho$ and (c) NOI versus the mean degree $z$ with fixed threshold $R=0.18$ [indicated by the vertical dotted line in (a)] and $\fracEasy = 0.2, 0.5, 1.0$. The lines are from theoretical calculation~\eqref{rho_equation}; the dots are from numerical simulations with $N=10^{6}$ nodes, averaged over $10^2$ realizations. The type of small-$z$ transition changes from continuous ($\fracEasy \in \{0.5, 1\}$) to discontinuous ($\fracEasy=0.2$). (c, inset) NOI at the small-$z$ transitions versus $\fracEasy$, displaying a peak at $\fracEasy_c\approx0.28$.}
\label{fig:critical_line}
\end{figure}

\section{Results}

We illustrate the main results with a simple yet rich case: an uncorrelated, two-layer (duplex) \ER network with identical mean degree $z$ in each layer. Also, each node has the same threshold $r_\alpha = R$ for both layers $\alpha \in \{1, 2\}$. Thus, $P(\mathbf{k})=P(k_{1})P(k_{2})$ with Poissonian degree distribution in each layer, and $Q(\mathbf{r})=\delta(r_{1}-R)\delta(r_{2}-R)$. Extending the formalism to more than two layers is straightforward and presented in part in~\cite{Brummitt2012}. (The degree distribution of correlated multiplex networks was introduced~\cite{KMLee2012,KMLeeBook}, and their robustness was recently studied~\cite{Min2014}.)

\subsection{Facilitating or inhibiting cascades}

First, we illustrate how changing the fraction of OR nodes, $\fracEasy$, either facilitates or inhibits global cascades and can affect the nature of the appearance of global cascades. We present the $(z, R)$-phase diagram displaying the regions of mean degree $z$ and threshold $R$ for which global cascades are likely and unlikely for various $\fracEasy$ [Fig.~\ref{fig:critical_line}(a)]. To obtain the boundary separating these parameter regions, we find local maxima of the number of iterations (NOI) of the recursion relation~\eqref{recursion}, a procedure comparable to examining the divergence of relaxation time at a phase transition in critical phenomena~\cite{Goldenfeld1992} and recently applied to cascading failures in interdependent networks \cite{Parshani2011}. This method more accurately locates the boundaries than the first-order cascade condition used in previous studies~\cite{Watts2002,Gleeson2007,Gleeson2008,Brummitt2012} because it accounts for nodes activated by more than one neighbor.

If $\fracEasy=1$, then all nodes follow the OR rule~\eqref{FOR}, which maximally facilitates global cascades compared to the single-layer case [compare the red boundary with the gray region in Fig.~\ref{fig:critical_line}(a)]~\cite{Brummitt2012}. One can also assess the effect of multiplexity by splitting a given network into multiple layers, which is also found to facilitate cascades for $\fracEasy=1$~\cite[Fig.\ 3]{Brummitt2012}.

As $\fracEasy$ decreases, more nodes follow the AND rule~\eqref{FAND}, which inhibits cascades and hence shrinks the cascade region [see the orange, green, purple, and blue boundaries in Fig.~\ref{fig:critical_line}(a)]. When $\fracEasy$ is less than approximately $0.3$, the cascade region becomes smaller than the single-layer case, showing that multiplexity can also impede cascades. If all nodes follow the AND rule (i.e., $\fracEasy = 0$), then global cascades are nearly impossible [see the dot-dashed boundary in Fig.~\ref{fig:critical_line}(a)].

\begin{figure}[t]
\includegraphics[width=.86\columnwidth]{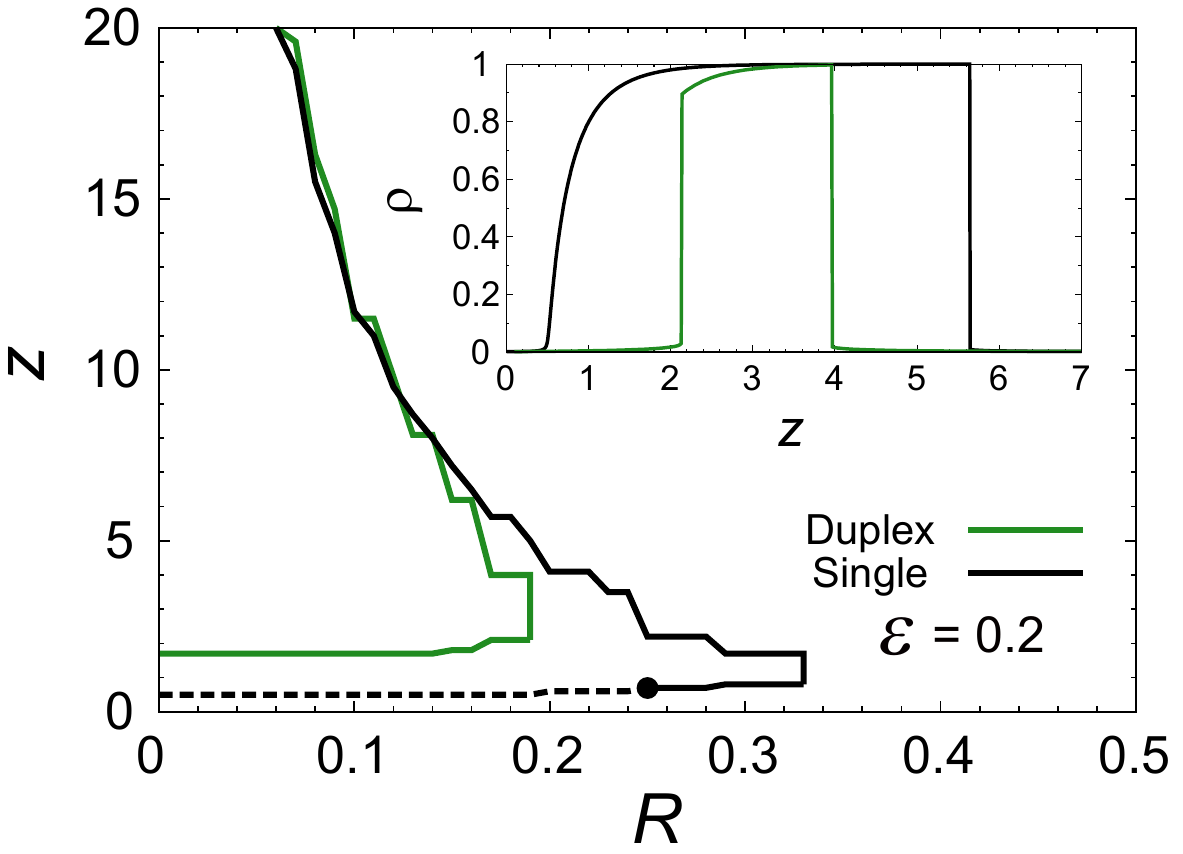} 
\caption{(Color online) The $(z,R)$-phase diagram of global cascades on duplex ER networks for the multiplex model  (green line) and the single-layer model that was designed to replicate the multiplex model (black line) with $\fracEasy=0.2$. The dashed and solid lines indicate the continuous and discontinuous transition, respectively. The black dot separates two different transition types. For the single-layer model, the mean degree is  $z_{\textrm{single}}=2z$, and the thresholds are given by $R/2$ for OR nodes and $R$ for AND nodes.  For the duplex model, the mean degree is $z_\text{duplex}=z$ in each layer, and the thresholds are given by $R$ for all nodes.}
\label{fig:compare} 
\end{figure}

Before proceeding further, we briefly address the issue of reducibility of multiplex dynamics to an equivalent dynamics on single-layer networks with appropriately chosen heterogeneous threshold distributions. In our example, an intuitive and reasonable choice would be to set the threshold equal to $R/2$ for a fraction $\fracEasy$
of nodes (to play the role of OR nodes) and equal to $R$ for the rest (to play the role of AND nodes) on a single-layer network with twice the mean degree, $z_{\textrm{single}}=2z$, as that of the layers of duplex network. The obtained ($z, R$)-phase diagram of this single-layer model is shown in comparison with that of the original multiplex model for $\fracEasy=0.2$, the case chosen deliberately to display discontinuous transitions in the multiplex model (Fig.~\ref{fig:compare}). The single-layer model differs from the multiplex dynamics not only quantitatively (different phase boundaries) but also qualitatively (different transition types). This result suggests that ``reducing'' the multiplex dynamics into a single-layer model should be highly nontrivial when the layer-dependent response in multiplex dynamics [such as Eqs.~\eqref{FOR} and~\eqref{FAND}] is not simply additive but nonlinear.

\begin{figure}[t]
\includegraphics[width=.92\columnwidth]{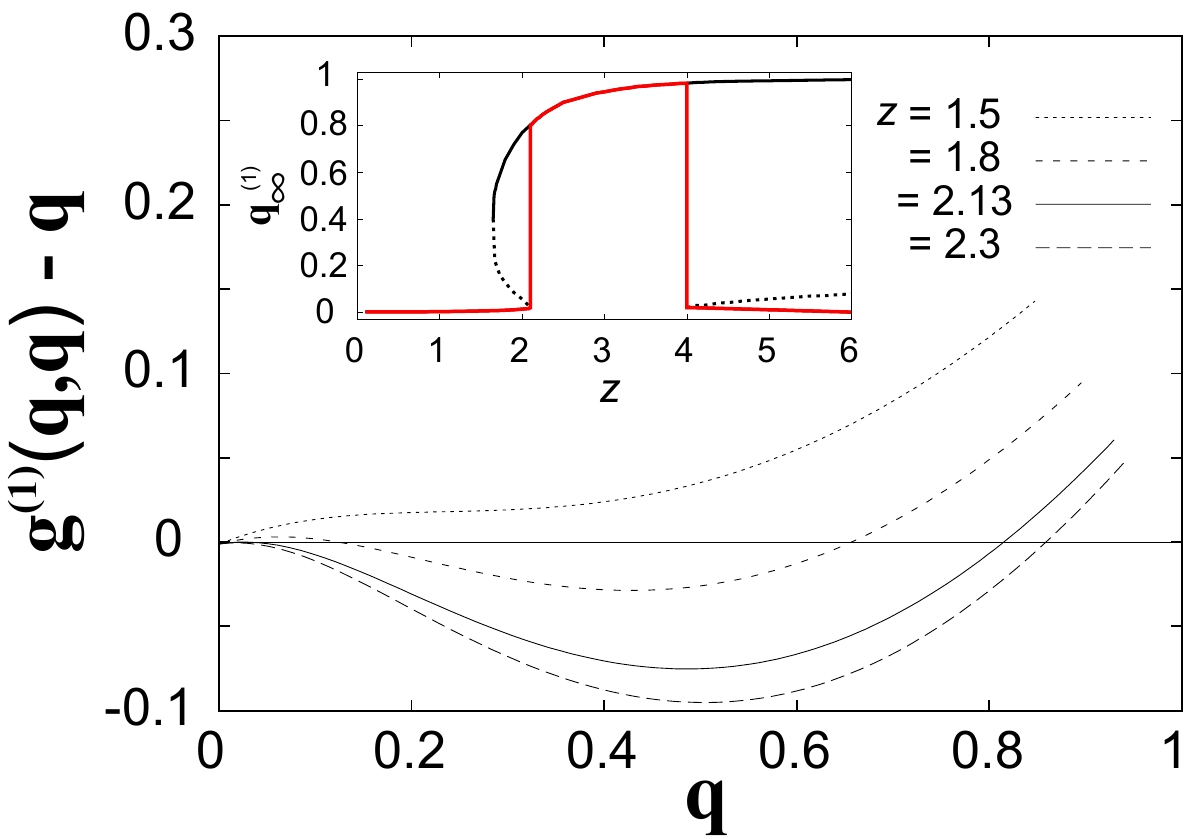} 
\caption{(Color online) (Main) Graphical solutions for the fixed points of the recursion~\eqref{recursion} for duplex ER networks with  $z=1.5, 1.8, 2.13, 2.3$ and $R=0.18$, for $\fracEasy=0.2$. (Inset) Bifurcation diagram of the roots of $\bf{g}(\bf{q})-\bf{q}$, with solid and dotted curves denoting the stable and unstable solutions, respectively. The red curve denotes the physical solutions for cascades starting from the small seed $\rho_0 = 10^{-3}$. Global cascades appear discontinuously at $z \approx 2.13$.
}
\label{fig:bifcurves}
\end{figure}

\subsection{Cusp catastrophe and tricritical-point scaling}

Not only does reducing the fraction $\fracEasy$ of OR nodes inhibit global cascades; it can also cause cascades to appear discontinuously. Previous work has shown that if every node in a single-layer network has the same threshold $R$, then the mean cascade size $\rho$ grows continuously and then drops discontinuously with increasing mean degree $z$~\cite{Gleeson2007}. In our multiplex model with a mixture of response rules,  
the small-$z$ transition for global cascade changes from continuous to discontinuous when $\fracEasy$ becomes sufficiently small [Fig.~\ref{fig:critical_line}(b)]. We find that the NOI for these small-$z$ transitions exhibits a peak at $\fracEasy_c \approx 0.28$ [Fig.~\ref{fig:critical_line}(c), inset], and later we show that this value is where the continuous transition becomes discontinuous.  In passing, we note that a discontinuous appearance of global cascades was also observed in a single-layer network with heterogeneous thresholds, displaying, however, a quite different phase diagram~\cite{Gleeson2007}. 

The bifurcation diagram (Fig.~\ref{fig:bifcurves}, inset) confirms this discontinuous transition. The fixed points $\mathbf{q}$ of recursion~\eqref{recursion} are roots of ${\bf{g}(\bf{q})-\bf{q}}=0$. In numerically simulated cascades, we observe only the smallest root (plotted as red in the inset of Fig.~\ref{fig:bifcurves}) because we consider small seed-sizes $\rho_0 \ll 1$. Bifurcation analysis of $q^{(1)} (=q^{(2)})$ reveals that the system undergoes a fold catastrophe: as $z$ increases from $0$, two new roots appear in a saddle-node bifurcation, 
and one of those roots (the unstable one) annihilates the small, stable root at another saddle-node bifurcation at 
the small-$z$ transition point, leaving only a large stable root (Fig.~\ref{fig:bifcurves}). Thus, as the network densifies, global cascades appear discontinuously. Increasing the fraction $\fracEasy$ of OR nodes beyond $\fracEasy_c\approx0.28$ eliminates the fold catastrophe, thereby restoring the familiar continuous transition~\cite{Watts2002,Gleeson2007}. 

\begin{figure}[t]
\includegraphics[width=.85\columnwidth]{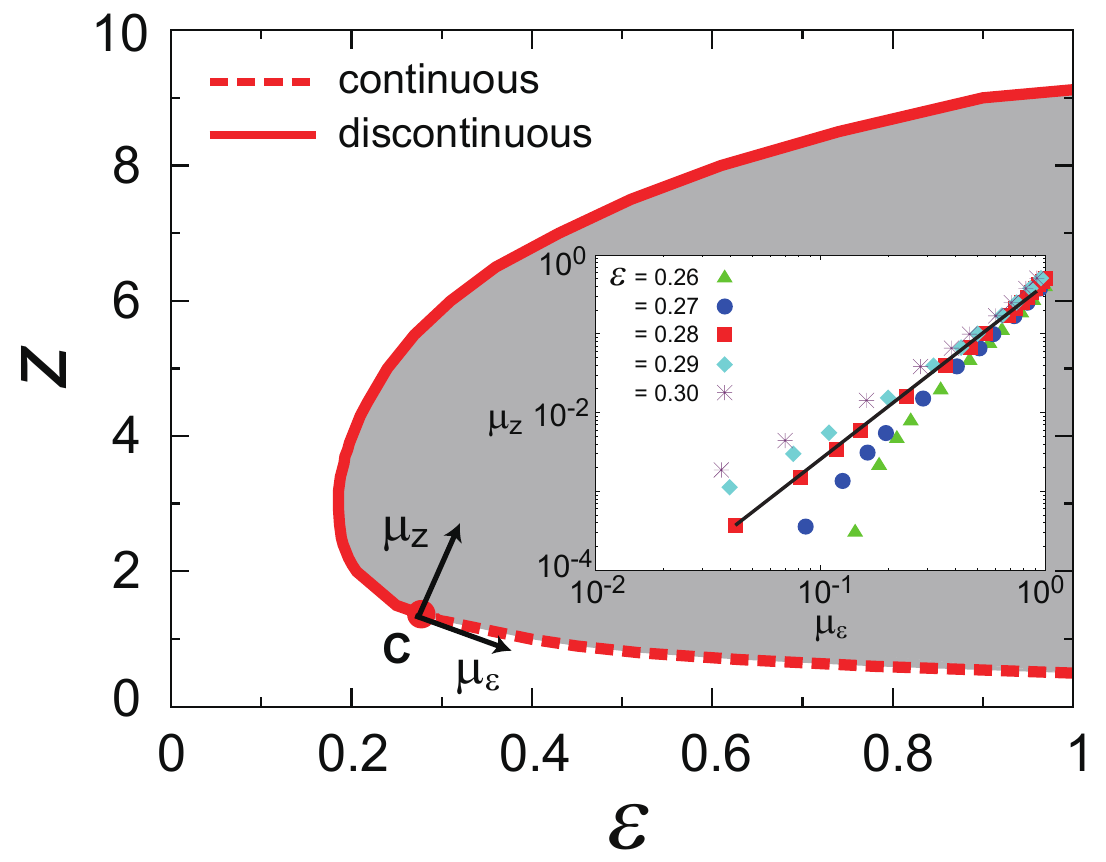} 
\caption{(Color online) (Main) The $(z,\fracEasy)$-phase diagram of global cascades of duplex ER networks with threshold $R=0.18$. The gray region indicates where global cascades occur. Continuous and discontinuous transitions occur along the dotted and solid curves, respectively, separated by the cusp point $\textrm{\bf C}=(\fracEasy_c, z_c) = (0.28, 1.36)$ marked by a dot. (Inset) The scaling relation along the continuous transition curve in the new coordinate system ($\mu_{\fracEasy}, \mu_{z}$) centered at a point on the transition curve for various $\fracEasy$. The best fit to the straight line is observed for the coordinate system centered at {\bf C}. 
The straight line has slope $2$, drawn as a guideline. 
}
\label{fig:mix}
\end{figure}

In short, the model undergoes a cusp catastrophe~\cite{Poston1996}, illustrated in Fig.~\ref{fig:mix}.  The cusp point $(\fracEasy_c, z_c)$ marks the parameters at which the line of continuous transitions and the line of discontinuous transitions join. This point can be associated with tricritical behavior when the initial seed size $\rho_0\ll1$: the continuous transition line undergoes scaling behavior as it crosses-over to the discontinuous one.
To describe the crossover behavior, one introduces two new variables $\mu_{\fracEasy}$ and $\mu_{z}$ (called ``scaling fields'') \cite{Riedel1972, Araujo2011, Cellai2011}, which are tangential and normal, respectively, to the continuous transition line. In the new coordinate system centered at the estimated cusp point $(\fracEasy_c, z_c) = (0.28, 1.36)$, the two scaling fields obey a power-law relation near the origin as $$\mu_{z} \sim \mu_{\fracEasy}^{1/\varphi_{t}},$$ with the crossover exponent $\varphi_{t} = 1/2$ (Fig.~\ref{fig:mix}, inset). Other choices for $(\fracEasy_c, z_c)$ were not compatible with the scaling.

\begin{figure}[t]
\centering
\includegraphics[width=.95\columnwidth]{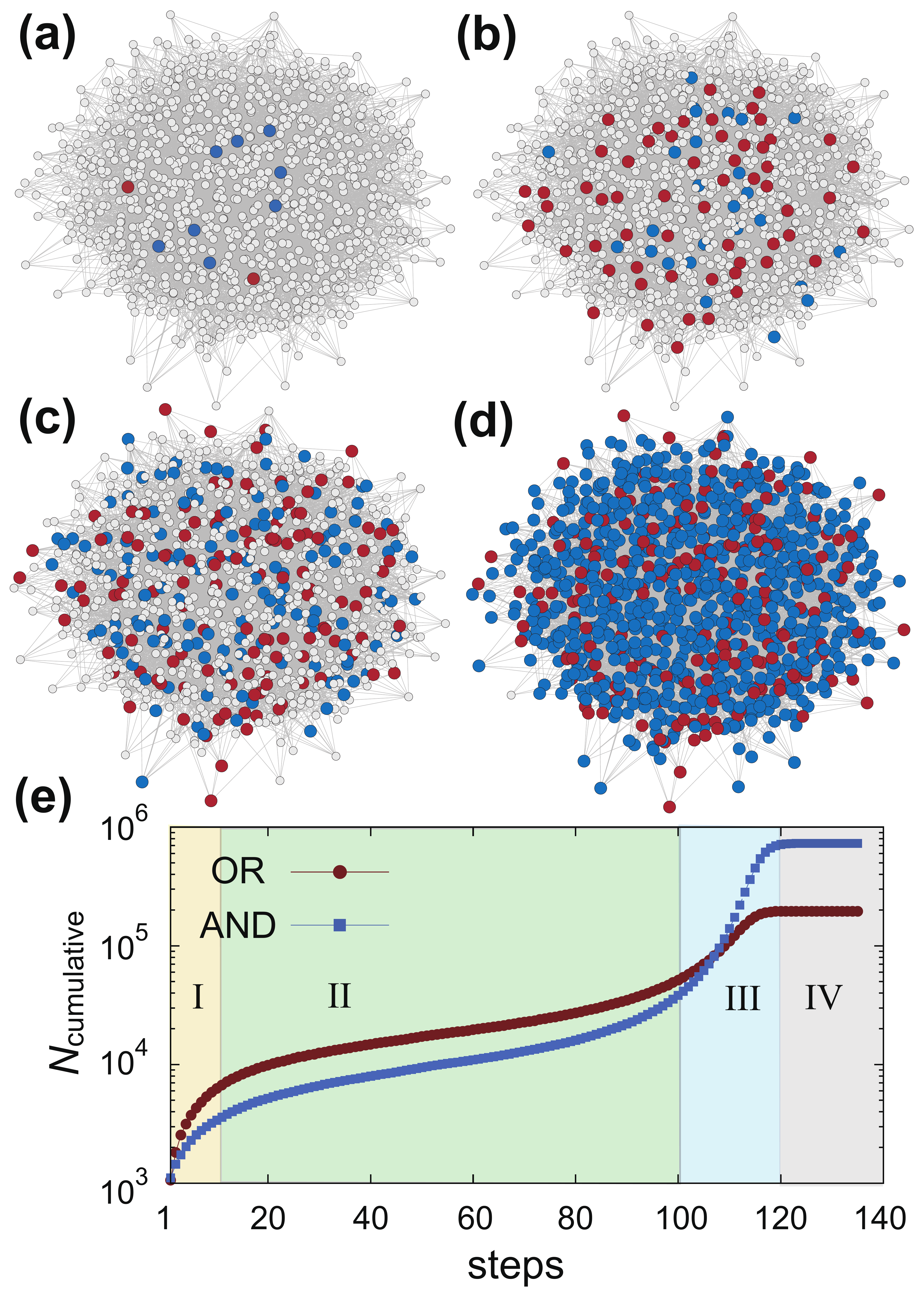} 
\caption{
(Color online) (a--d) Example cascade snapshots in a small duplex ER network of $N=10^3$ nodes with $\fracEasy=0.2$. The red and blue nodes are active nodes that follow the OR and AND rules, respectively. 
Starting from a random 1\% of seed nodes (a), the OR nodes activate in greater numbers in the first few steps (b). As time proceeds, however, AND nodes activate in greater numbers (c) and eventually dominate (d). 
(e) Cumulative numbers of active OR and AND nodes in a numerically-simulated cascade on a duplex ER network of  $N=10^6$ nodes with $z=2.3$, just above the small-$z$ transition point for $\fracEasy = 0.2$. Initial exponential growth  (I, yellow)  is rapidly slowed due to the ``stubborn'' AND nodes (II, green), but once enough nodes are active the activation resumes the exponential growth, and AND nodes overtake OR nodes (III, blue), finally reaching saturation (IV, gray).}
\label{fig:step}
\end{figure}

\subsection{Slowed cascades near the cusp point}

Response heterogeneity affects not only whether cascades appear discontinuously; it also affects who activates when and how slowly the cascade progresses. As depicted in Fig.~\ref{fig:step}, a typical global cascade near the cusp point can be qualitatively divided into four stages [labeled I--IV in Fig.~\ref{fig:step}(e)]. Initially, activation grows exponentially, and the more susceptible OR nodes activate in greater numbers than the AND nodes, even though OR nodes are less numerous (because $\fracEasy= 0.2$ in Fig.~\ref{fig:step}). In stage II, the rates of activation slow for both types of nodes. What is particularly interesting about stage II is that the presence of AND nodes significantly delays the global cascade. If the goal is to prevent large cascades (as in bank regulation), then stage II provides a crucial window of opportunity for intervention. After sufficiently many nodes have activated, the AND nodes activate at a faster rate and eventually overtake the number of active OR nodes (stage III), as there are more AND nodes in the network with $\fracEasy = 0.2$. Finally, the activations saturate for a finite system (stage IV).

\section{Summary and Discussion}
Introducing new network layers, such as adding new social media or creating novel ways of lending, can facilitate or impede threshold-driven cascades, depending on how nodes respond to their multiplex surroundings (Fig.~\ref{fig:critical_line}). For networks in which most nodes can be activated through any one of the layers (i.e., for large $\fracEasy$), global cascades become likely---even for networks that would have been too dense to allow global cascades if there were just one channel of influence. By contrast, if most nodes wait to activate until their thresholds are met in each and every layer (i.e., if $\fracEasy$ is small), then global cascades occur rarely, if at all. 
However, when global cascades do occur in this small-$\fracEasy$ regime, they appear discontinuously (and at a larger network density) as the network densifies (Figs.~\ref{fig:critical_line} and \ref{fig:mix}). At the same time, such discontinuous global cascades take considerably longer to develop (Figs.~2 and 5). 
The AND response rule in our model [Eq.~\eqref{FAND}] is analogous to the rule of mutual connectivity in mutual percolation on multiplex networks~\cite{Buldyrev2010}. Therefore, the discontinuous transition observed in our model shares a similar origin with that in mutual percolation--type problems in interdependent and multiplex networks~\cite{Buldyrev2010,Gao2011,Baxter2012,Zhou2014,Viability,Baxter2014}.

Real multiplex complex systems such as social and financial systems have considerably greater structural and dynamical complexity than the model studied here, including, for example, interlayer correlations \cite{KMLee2012,Min2014}, link overlap \cite{Overlap}, heterogeneous network structure \cite{Watts2002}, heterogeneous thresholds \cite{Gleeson2007}, and different types of cascading failure dynamics on weighted networks~\cite{Mirzasoleiman2011}, to name only a few. Regarding the multiplex response, the two simplified response rules studied in this work could be made more realistic by considering, for example, different combinatorial response rules and/or time-dependent, adaptive response rules \cite{CAS}. We hope that our study of a simple theoretical model can aid in stimulating extensions that better capture real cascading phenomena.

\begin{acknowledgments}
This work was supported in part by Basic Science Research Program (No.\ 2011-0014191) through NRF grants funded by the MSIP; CDB by NSF EAPSI, Award 1107689, the DTRA Basic Research Award HDTRA1-10-1-0088, and the DoD through the NDSEG Program.
\end{acknowledgments}

\end{document}